\documentclass[11pt,twoside]{article}
\usepackage{asp2006}
\usepackage{epsf}
\usepackage{graphics}
\usepackage{lscape}
\def\arg#1{{\it#1\/}}
\let\prog=\arg
\def\edcomment#1{\iffalse\marginpar{\raggedright\sl#1\/}\else\relax\fi}
\reversemarginpar

\begin{document}
\title{The Signature of Flares in VIRGO Total Solar Irradiance Measurements}
\author{A. Quesnel$^1$, B.~R, Dennis$^2$, B. Fleck$^3$, C. Fr{\" o}hlich$^4$, H.~S. Hudson$^5$,
and A.~K. Tolbert$^2$}
\affil{$^1$Ecole Polytechnique Palaiseau, $^2$NASA/GSFC, $^3$ESA, $^4$PMOD/WRC Davos, $^5$UC Berkeley}

\begin{abstract}
We use Total Solar Irradiance (TSI) measurements from the VIRGO (Variability of solar
IRradiance and Gravity Oscillations) instrument on board SOHO to obtain preliminary estimates
of the mean total radiative energy emitted by X-class solar flares. 
The basic tool is that of summed-epoch analysis, which has also enabled us
to detect and partially characterize systematic errors present in the basic data. 
We describe these errors, which significantly degrade the photometry at high frequencies.
We find the ratio of GOES 1-8\AA~luminosity to total bolometric luminosity to be of order~0.01.
\end{abstract}

\vspace{-0.5cm}
\section{Introduction}

The detection of solar flares against the glare of the photosphere has only previously been possible via imaging, whereas many kinds of \textit{stellar} flares can be readily detected via broad-band photometry
of integrated starlight.
This is often because of a cooler, fainter photospheric background, as in the classical dMe flare stars.
The complete energy distribution of a solar flare across many wavebands (especially, for example the vacuum UV) has also remained poorly understood \citep[e.g.][]{emslie}.
Another problem for characterizing the  bolometric luminosity of a solar flare lies in the photospheric fluctuations, mainly a broad continuum of power due to granulation \citep[e.g.][]{hudson} and having
an rms amplitude of roughly 50~ppm in the 2-8~mHz band.
The Total Solar Irradiance (TSI) instruments flown in space since the late 1970s typically sample this band.
Flare light curves at various wavelengths suggest that this band contains much of the flare power,
although variations well above 1~Hz do occur \citep[e.g.][]{dennis}.

Only recently has it become possible to detect solar flares bolometrically \citep{woods}, and only
very convincingly for a single event, the X17~flare of 2003 October~28.
The solar background noise level corresponds to 50~$\mu$mag in astronomical terms, and it is remarkable that this fluctuation obscures most solar flares, whereas stellar flares often have
increases exceeding one full magnitude.
Kretzschmar (2008)
\nocite{kretzsch}
has introduced the technique of superposed epoch analysis to detect multiple flares in the mean, and we adopt this scheme for the analysis presented here.
We also study TSI observations from the VIRGO instrument on board SOHO, but the same techniques should work well for other instruments.
Our analysis has revealed sources of unwanted variance in the TSI signal, which we intend to correct in a future analysis.

\section{The data}

The VIRGO instrument consists of three types of detector: the absolute radiometers DIARAD and PMO6V, plus a set of ``sun photometers'' with broad-band blue, green, and red responses.
The analysis reported here uses only the PMO6V data.
These are electrical substitution measurements on the solar signal observed in a cavity with high absorptivity, utilizing feedback to maintain uniform conditions during a one-minute shutter cycle \citep{frohlich}.
The spectral response of such an instrument includes all energetically significant flare wavelengths.
We use the PMO6V level-2 data. 
These have been converted to physical units and contain all the corrections known \textit{a priori} for instrument-related systematic effects. 
The level-2 data also contain corrections for degradation, which were calculated from an \textit{a posteriori} analysis developed and agreed upon by the science team.
Figure~1 (left) shows the noise properties of the data for the year 2003, as rms fluctuations for daily intervals with and without high-pass filtering at 1~mHz.

Despite their high processing level, we have found that the data still contain unwanted sources of variance.
Figure~1 (right) illustrates an ``odd-even'' effect discovered in our superposed-epoch tests; consecutive data points systematically differ by $\sim$35~ppm.
This artifact appeared in the time series of PMO6V data when the LOI instrument was switched on, 
but we do not yet understand it.
Another LOI-related artifact can be seen in power spectral analysis not shown in this paper.
It is broad-band in nature and dominates the spectrum between about 7~mHz and the Nyquist frequency (8.33~mHz); unfortunately most  of the solar-flare variability occurs in this band.

\begin{figure}
\plottwo{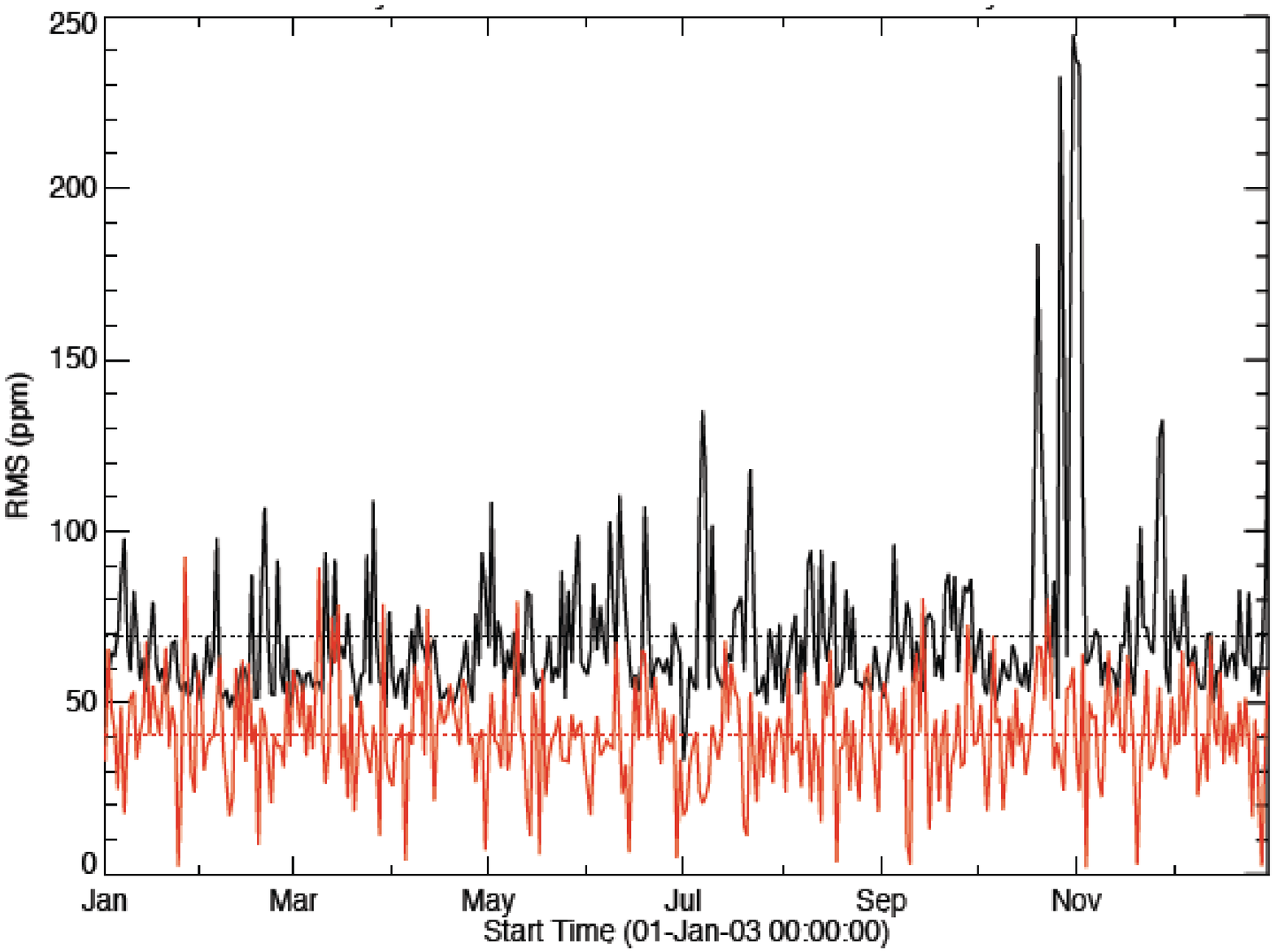}{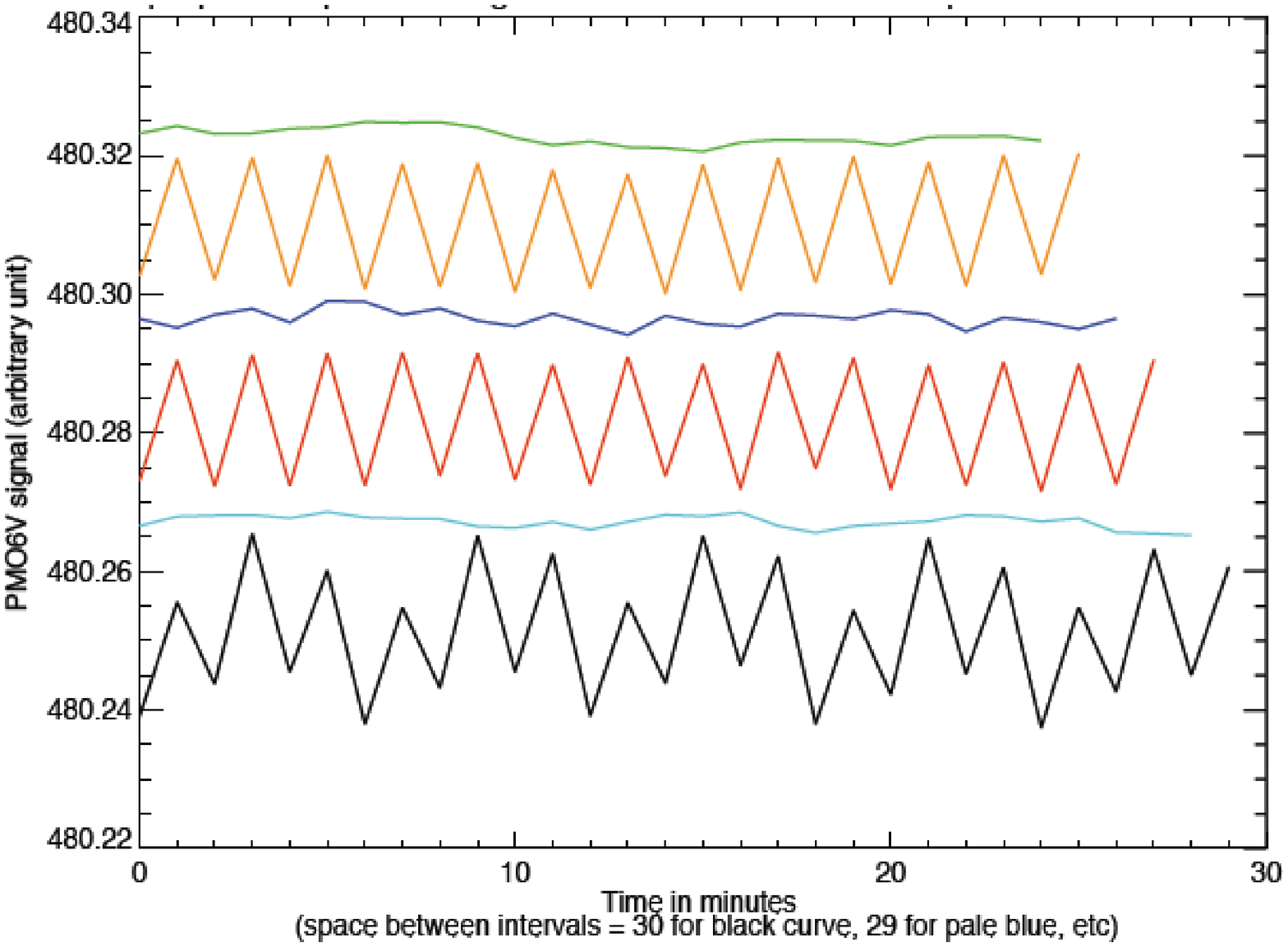}
\caption{
\textit{Left:} RMS fluctuations for one-day intervals during 2003.
The upper time series shows unfiltered data, in which sunspot and other TSI effects produce obvious spikes.
The lower curve shows the same data high-pass filtered at 1~mHz.
\textit{Right:} illustration of the ``odd-even'' effect -- the lines show superposed epochs with odd and even numbers of samples, spaced vertically for clarity.
}
\end{figure}

\section{Summing on GOES epochs}
Following Kretzschmar (2008) we use the GOES soft X-ray observations as a ready source of ``key times'' for the flare summed-epoch analysis.
This technique was refined by Charles Chree \citep{chapman}; it is a powerful technique, but because the co-adding of data beats the noise down, if often reveals artifacts that can be misconstrued.
In Figure~2 we show two summed-epoch analyses of the PMO6V data against the GOES data: 
one that uses the 1-8\AA~maximum time as the epoch, and one that uses the peak of its derivative.
This follows the  \cite{neupert}  effect, in which this derivative tends to match the flare impulsive 
phase.

\begin{figure}
\plottwo{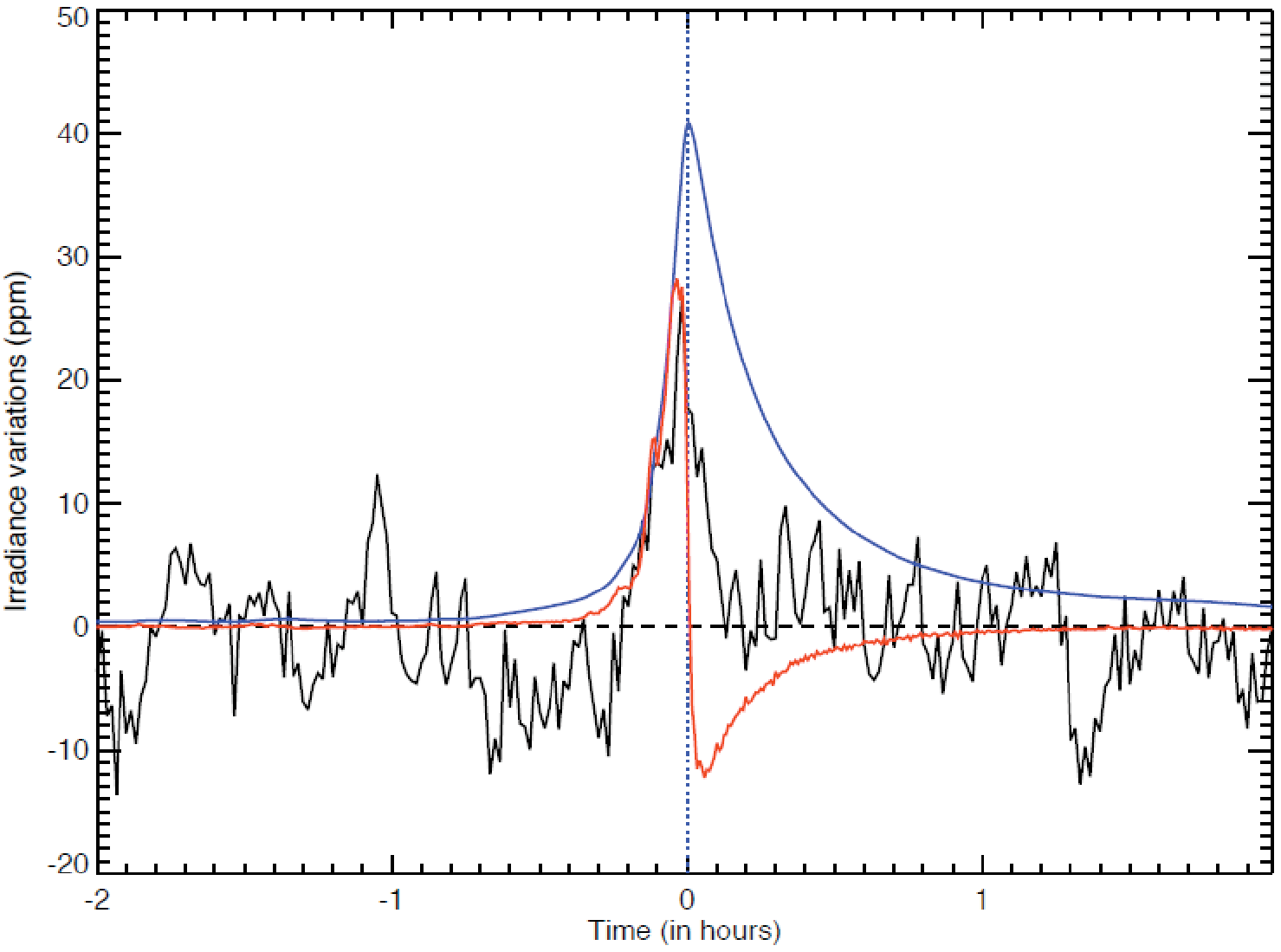}{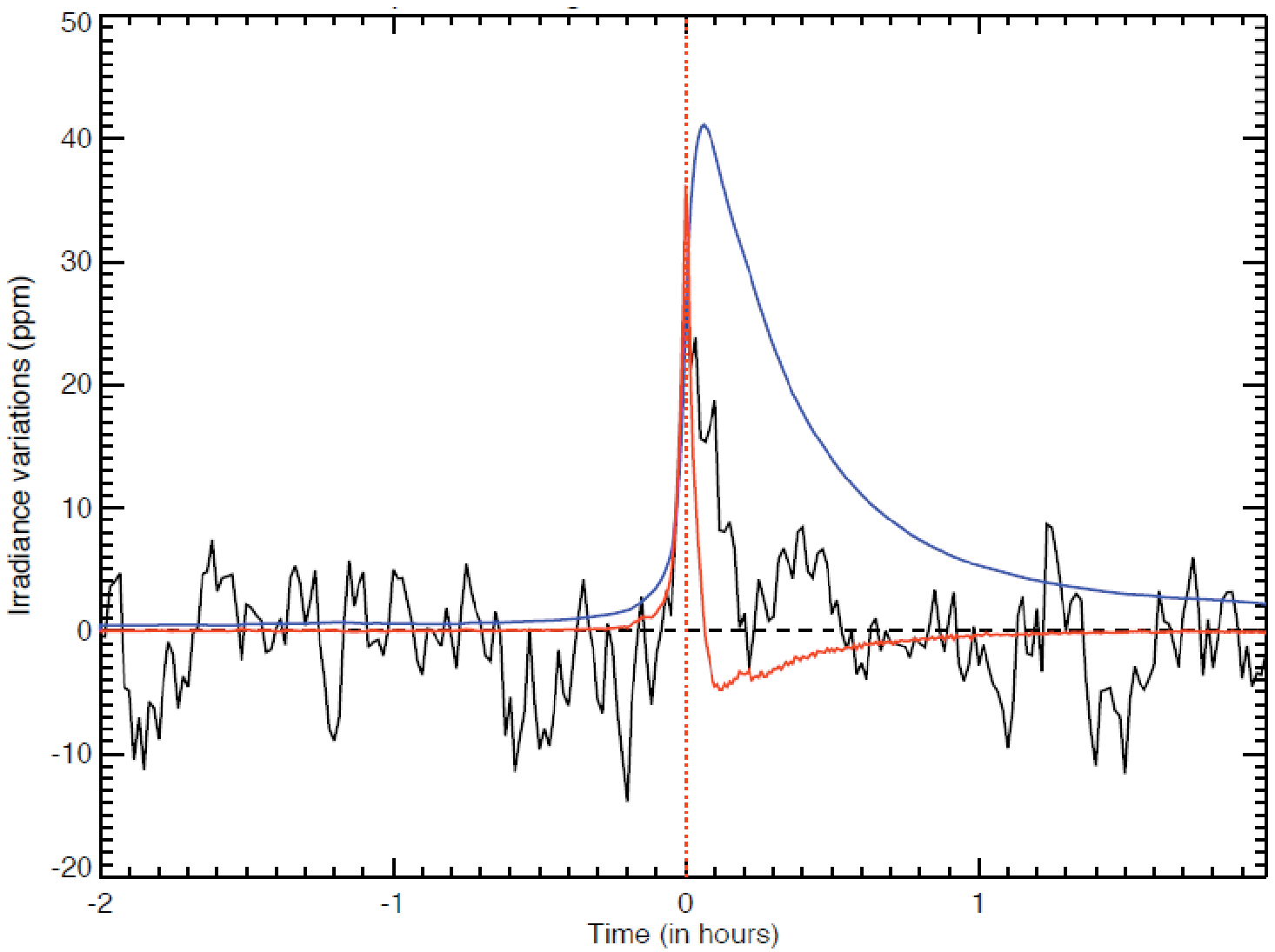}
\caption{
\textit{Left:} summed-epoch analysis using GOES 1-8\AA~peak times, for 4-hour intervals.
\textit{Right:} summed-epoch analysis using the derivative of the GOES time series.
Both analyses use all 117~X-class flares occurring during the SOHO lifetime from 1996 through
to the time of writing.
}
\end{figure}

The comparison of the two analyses in Figure~2 suggests that the impulsive phase of the flare,
as marked by the GOES time derivative, organizes the TSI signals better than the GOES maximum time.
We interpret this to mean that the white-light flare and its associated UV~continuum emission has a major energetic significance, as hinted at by the time series of the 2003 October~28 flare detected in the SORCE TSI data by \cite{woods}.

\section{Future Improvements}

We discuss here a few possible improvements in the analysis described here. 
The random noise in the PMO6V data, and in other TSI measurements, may actually not contribute much of the variance seen in Figure~1.
If so, the TSI fluctuations during an individual flare have a structure that may be well defined by an
independent data set.
For example, we speculate that these variations outside flare times may well correlate quite 
exactly with the integrated signals from the whole-Sun images to be obtained from the Solar Dynamics Observatory.
This correlation could then be used to correct the TSI data and thereby obtain substantially improved photometry.
Other gap-filling techniques may also be possible.
TSI observations with higher time resolution (unfortunately, none exist at present) would remove the aliased broad-band background fluctuations.
Finally, the analysis we describe here is only a preliminary one, and we plan in the future to repeat the analysis after correcting for the artifacts shown in Figure~1.

A summed-epoch analysis in general should improve the signal-to-noise ratio for events of the same magnitude, but flares have a distribution in total energy that follows a power law \citep{hudson91,crosby}.
Including weaker flares to a randomly-selected set might add more noise than signal, depending upon the power-law index of the distribution.
We have analyzed this effect and will discuss it in our full paper on the flare analysis of the PMO6V data.

\section{Conclusions}

We have reported initial results of summed-epoch analysis for the total irradiance of solar flares.
This analysis suggests a strong correlation with the impulsive phase of a flare, identified here by the
time derivative of the soft X-ray flux.
We use the analysis to obtain an estimate of $L_X/L_{bol}$, which can then be compared with similar estimates for different kinds of stellar flares, for which parameters such as gravity, rotation, or composition can be studied.
Our analysis, that of Kretzschmar (2008), and the synthesis of Emslie et al. (2005) all agree that this ratio is of order~0.01, using the GOES 1-8\AA~passband.
We have also found a source of unwanted noise in the observations, which we plan to correct in a future more
sensitive analysis of these data.

\bigskip\noindent
{\bf Acknowledgements:}
SOHO is a project of international cooperation between ESA and NASA.
HSH thanks NASA for support under grant NNX07AH74G.

{}

\end{document}